\documentclass[12pt]{article}
\usepackage[english]{babel}

\usepackage{amsmath}
\usepackage{graphicx}
\usepackage[colorinlistoftodos]{todonotes}

\usepackage[font=scriptsize,labelfont=bf]{caption}
\usepackage{amsmath}

\usepackage{graphicx}
\usepackage{xcolor}
\usepackage{amssymb}
\usepackage{enumitem}
\usepackage{mathtools}

\usepackage{amsfonts}
\usepackage{amssymb}
\usepackage{amsbsy}
\usepackage{pifont}
\usepackage{verbatim}

\usepackage{hyperref}
\hypersetup{
    colorlinks,
    citecolor=black,
    filecolor=black,
    linkcolor=black,
    urlcolor=black
}
\usepackage{graphicx}
\graphicspath{ {images/} }

\usepackage{subcaption}
\usepackage{times}
\usepackage{array,multirow}
\usepackage{xspace}
\usepackage{xcolor}
\usepackage{color, colortbl}
\usepackage{balance}
\usepackage{paralist}
\usepackage{amsmath}
\usepackage{amsfonts,amssymb,amsthm}
\usepackage{graphicx}
\usepackage{pifont}
\usepackage{array}
\usepackage{booktabs}
\usepackage{tikz}
\usepackage{bm}
\usepackage{enumitem}
\usepackage{url}

\usepackage[
backend=biber,
style=numeric,
sorting=none
]{biblatex}

\renewbibmacro{in:}{}

\usepackage[T1]{fontenc}

\newcommand\definition[2]{\ding{118}\xspace \textsf{#1}\xspace$\bm{\rightarrow}$\xspace(#2):\xspace}

\newcommand\elgamal{El-Gamal\xspace}
\newcommand{\etal}{\textit{et al.}\@\xspace}
\newcommand\hashtopoint{H\xspace}

\begin{document} 

\begin{titlepage}

\newcommand{\HRule}{\rule{\linewidth}{0.5mm}} 

\center 
 

\textsc{\LARGE University College London}\\[1.5cm] 
\textsc{\Large Computer Science Department}\\[0.5cm] 
\textsc{\large A masters dissertation\footnote{This report is submitted as part requirement for the MSc in Information Security at University College London. It is substantially the result of my own work except where explicitly indicated in the text. The report may be freely copied and distributed \textit{provided} the source is explicitly acknowledged.} on}\\[0.5cm] 


\HRule \\[0.4cm]
{ \LARGE \bfseries Coconut E-Petition Implementation}\\[0.4cm] 
\HRule \\[1.5cm]
 

\begin{minipage}{0.4\textwidth}
\begin{flushleft} \large
\emph{Author:}\\
Jad \textsc{Wahab} 
\end{flushleft}
\end{minipage}
~
\begin{minipage}{0.4\textwidth}
\begin{flushright} \large
\emph{Supervisors:} \\
Prof. George \textsc{Danezis} \\
\small{Alberto Sonnino \\
Mustafa Al Bassam}
\end{flushright}
\end{minipage}\\[2cm]



{\large \today}\\[2cm] 


\vfill 

\end{titlepage}

\addtocounter{page}{1} 


\setcounter{secnumdepth}{0} 

\begin{abstract}

In this dissertation project, we describe and implement a practical system application based on a selective disclosure credential scheme, namely the Coconut credential scheme\cite{sonnino_coconut:_2018}. The specific application here is an electronic petition system with the distinctive added feature of unlinkability as well as anonymity: such that no information about the anonymous petition voter is linkable back to the individual. In other words, there is no data leaked about who voted in the petition, just that the users who did, were indeed eligible and authorized to vote. As for the implementation, the client-side is done using JavaScript so that the client can trustlessly compute the cryptographic constructions individually, whereas the server-side is done using Node.js, but can easily be replaced by a more sophisticated and secure structure such as a permissionless blockchain platform.

\end{abstract}

\clearpage

\renewcommand{\abstractname}{Acknowledgements}
\begin{abstract}
I would like to thank my supervisors Professor George Danezis, Mustafa Al Bassam, and especially Alberto Sonnino for their support and guidance towards the completion of this dissertation project.
\par

I would also like to thank my friend Joe Haddad who helped me with troubleshooting some programming issues I had. Moreover, I would like to thank Jedrzej Stuczynsk for his contributions and readied support if I ever needed help.

\end{abstract}
\clearpage

\tableofcontents
\clearpage

\section{Introduction}

\subsection{Motivation and Goal}
With the recent increased interest and research in applied cryptography and privacy enhancing cryptographic techniques, there has been significant innovation, however with the implementations slightly lagging behind. After learning about these findings and techniques over the year, I decided to create a specific implementation which relied on a cryptographic approach, namely a selective disclosure credential system called Coconut \cite{sonnino_coconut:_2018}. Selective disclosure credentials \cite{cl} \cite{chase_algebraic_2014}
allow issuance of a specific credential (having one or more attributes) to a user with the ability to unlinkably reveal or "show" said credential at a later instance, for purposes of authentication or authorization. The system provides the user the ability to also "show" specific attributes of that credential or a specific \emph{function} of the attributes embedded in the credential; e.g. if the user has an attribute $x$ representing their age, let's say $x=21$, they can show that $f(x) > 21$ without revealing $x$.

\par
One of the benefits from this approach is that users can be authenticated based on a credential while still preserving their anonymity. In other words, the user can be authenticated but the credential cannot be linked back to the user. Specifically, this project covers the application of an electronic petition system build on top of the Coconut credential system. There are a number of electronic petition systems that are currently being developed, but none that proved to be really complete or ready to be used in practical, large scale conditions. There were some implementations that try to make e-petition systems more secure and anonymous through the use of blockchain systems\cite{kiayias_smart_2017}\cite{shah_block_nodate}\cite{votewatcher}\cite{followmyvote} however they were still missing the feature of unlinkability, which is how this system is more encompassing. The motivation behind needing this added feature is to help prevent petitions from being vulnerable to corruption or interference and help provide a fairer result for the participants. As seen with the situation and events happening in Barcelona, \cite{stothard_barcelona:_2018} such a censorship-resistant feature would definitely be useful when authorities "seize[d] ballot boxes" \cite{noauthor_barcelona_nodate} ahead of the Catalan referendum. Hiding the metadata regarding \textit{who} voted for whom would definitely be useful in such a scenario. The only data which would be leaked \textit{if} at least one authority  maliciously leaks it, is the data concerning who is \textit{eligible} to vote, as opposed to who \textit{actually} voted. This should not happen in the case where all authorities are honest. Also, unlinkability makes the system more secure and private, even in the case that all voters collude to try and breach a voter's privacy (ie. perfect ballot secrecy). This is because even if the vote is breached, there is no information about who this specific vote belonged to in the first place.

\subsection{Organization/Structure of the Review}
This dissertation project revolves around the implementation of an e-petition scheme based on the Coconut selective credential disclosure system and an additive ElGamal homomorphic/threshold decryption system. We first go over a background of alternative methods for e-petition schemes as well as previous implementation of selective disclosure credential system, Tangerine \cite{stuczynski_tangerine:_nodate}. 
Following that, we introduce a high-level overview of the coconut architecture and go over a few notations and assumptions. Next, we go through the Coconut system definitions and design coupled with an additive homomorphic encryption scheme for petition voting. Afterwards, we describe the system implementation of the e-petition in parallel with the system design proposed. The system implementation mainly consists of a fully functional prototype in Javascript mainly so that it is completely trustless and client/front-end is able to do their own cryptographic computations in the browser. The back end is done in node.js but can very simply be replaced by a more secure back-end such as a blockchain platform. Then, we go over some testing and evaluation of the system implementation. Finally, we conclude the report and touch upon future directions and work to be done.

\clearpage
\section{Background and Related Work} \label{bg}
\subsection{Alternate E-Petition Implementations}
Traditional paper-based, physical voting systems are clearly not scalable and inherently incorporate numerous centralized points of failure and limitations. For example, you must trust numerous entities in the system not to leak data about you, plus you must trust that the people in charge do not do anything malicious to interfere with the petition or even prevent or censor people from voting (as in Barcelona \cite{noauthor_barcelona_nodate}), etc. However, a transition to digital voting systems undoubtedly opens the doors to significant security vulnerabilities, risks, and limitations. \par

There have been many such applications in the past, however as of the time of writing, there have not been which confidently solve the critical issues of at least security and privacy. Some attempts, such as the Virginia WINVote machine in 2015\cite{noauthor_virginia_nodate}, have even failed miserably and ended up being decommissioned as a result of the security assessment \cite{security-assessment} done which embarrassingly highlighted all of the security flaws and vulnerabilities of the system. 
\par

Votebook\cite{votebook}, a permission blockchain system for voting implemented by New York University, was an attempt to utilize the security of a blockchain system that resembles the current physical system with voting booths/stations and paper results that actually back the digital system on the blockchain. Their approach is a conservative one and favors a trade-off of security and integrity over privacy and censorship-resistance. Of course this is definitely legitimate approach in terms of security and reliability, however in some cases, such as with the Barcelona situation mentioned above, \cite{stothard_barcelona:_2018} privacy and censorship-resistance is crucial for the petition; especially when the police are "seiz[ing] ballot boxes" \cite{noauthor_barcelona_nodate} and trying to influence and censor the petition itself.
\par

There have been numerous other electronic voting and petition schemes/systems which are trying to leverage the blockchain security guarantees. However, as of the time of writing, to my knowledge, most feedback has been very skeptical with none of these systems actually confidently being used in practice. Even though some systems claim to have "the world's first secure open-source online voting software," \cite{followmyvote} in my opinion, there are still many issues in terms of security, privacy, censorship, fairness and scalability which are yet to be solved.
\par

Some of the current projects could actually benefit from adding a private credential scheme like Coconut to provide unlinkability for the users/voters. For example, the Ethereum implementation of the Open Vote Network protocol \cite{kiayias_smart_2017} could benefit from adding an extra credential step at the beginning of the system. The Open Vote Network protocol is neat protocol where the tally of votes can be computed by anyone. Basically, this is done by having all voting parties compute emphemeral keys based on each others public keys and use that to cast their vote. Once they all cast their vote to the blockchain (which is used essentially as a public immutable bulletin board), anyone (even a mere observer) can just compute the tally of votes. This is done by multiplying all of the cast votes together and because of the way the ephemeral keys are computed, the encryption cancels out. Then, to get the actual tally number from that value (which is basically $g^{\sum v_i}$ where the $v_i$'s are the different votes),  the discrete logarithm must be done using an exhaustive search of possible outcomes (since the result is relatively small this can be done).
\par

\subsection{Tangerine Signature System Criticism} \label{tang}
The Tangerine Signature System is a multi-authority selective disclosure credential system that was previously used to implement an electronic cash system done by a fellow UCL student, Jedrzej Stuczynski \cite{stuczynski_tangerine:_nodate}. Even though the Tangerine system is similar to the Coconut credential system \cite{sonnino_coconut:_2018}, it exhibits two major flaws/limitations in terms of the anonymity and unlinkability security properties. The first of them, relating to anonymity, being the fact that there is no blinding of the secret value in the commitment. This is a vulnerability in terms of security as the secret can be compromised by and active adversary. Also, if two users request credentials on respectively the values $m_1$ and $m_2$ such that $m_1=m_2$, the issuing authorities learn that these two users requested credentials on the same attribute by simply observing $C_{m_1}=g_1^{m_1}=g_1^{m_2}=C_{m_2}$; this breaks blind issuance. This may not be such a critical problem in some cases (such as the e-petition) where the credential is a random value but in other cases where the credential is someone's age or name for example, it definitely is. To solve this problem, Coconut uses Pedersen commitments which are \emph{unconditionally hiding}, as done in the \textsf{PrepareBlindSign} step of the Coconut implementation, namely replacing the commitment $c_m = g_1^m$ by a blinded commitment $c_m = g_1^m h_1^o$. \par 

The second vulnerability, concerning unlinkability, is the fact that in Tangerine's \textsf{prepareForVerify} step, matching \textsf{showBlindSign} or \textsf{ProveCred} of Coconut (depending on which version of the Coconut paper used), does not include any randomness in the creation of $\kappa$. This is fundamental in preserving the user's unlinkability against an active or even passive adversary. The attack is similar to the one against blindness where if two users have the same credentials on respectively the values $m_1$ and $m_2$ such that $m_1=m_2$, they would have the same $\kappa_{m_1}=\alpha\beta^{m_1}=\alpha\beta^{m_2}=\kappa_{m_2}$. The solution implemented in Coconut is adding randomness to change $\kappa$ from $\kappa=\alpha\beta^m$ to $\kappa=\alpha\beta^m g_2^r$. However, in order to keep the bilinear pairing valid, another component $\nu=\left(h'\right)^r$ had to be added, which doesn't pose any added non-negligible difficulties.

\subsection{High-level overview of Coconut architecture}
Coconut is ``a selective disclosure credential system, supporting threshold credential issuance of public and private attributes, re-randomization of credentials to support multiple unlinkable revelations, and the ability to selectively disclose a subset of attributes''\cite{sonnino_coconut:_2018}. The scheme is an extension of Pointcheval and Sander's work, \cite{pointcheval_short_2015} which is essentially a more efficient construction with the same properties of CL-signatures \cite{cl}. The main contribution of the Coconut scheme is that it introduces threshold issuance.

    \begin{figure}[h]
        \centering
        \includegraphics[width=.35\textwidth]{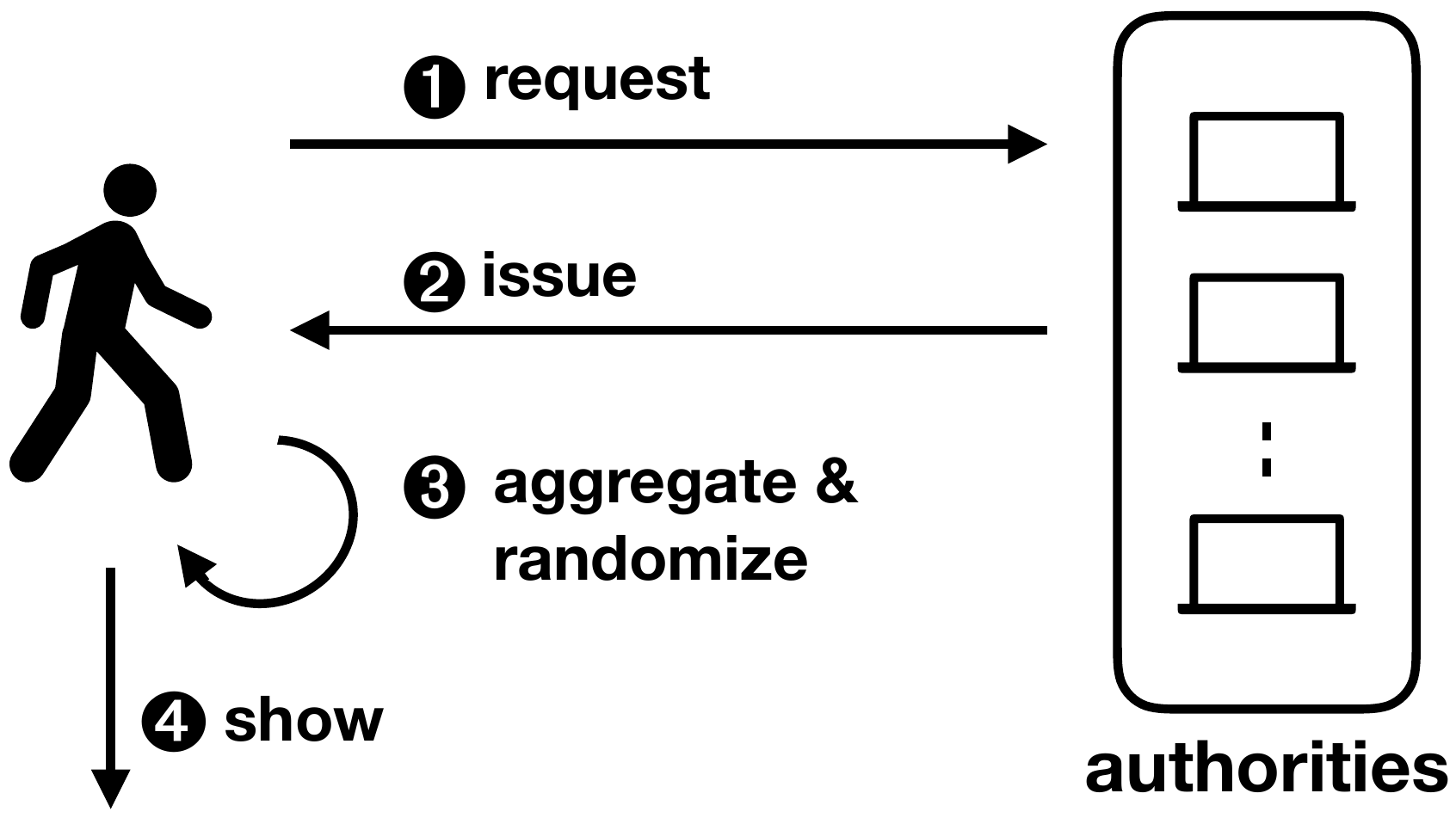}
        \caption{A high-level overview of the Coconut architecture, taken from Coconut paper \cite{sonnino_coconut:_2018}}
        \label{fig:coco-arch}
    \end{figure}

As illustrated in figure \ref{fig:coco-arch} above, in the $1^{st}$ step the user requests a credential from a set of authorities. In the $2^{nd}$ step, 
each authority issues the user with a partial credential which the user aggregates into a full credential (when they have a threshold number of shares) and randomizes in the $3^{rd}$ step. Finally, in the $4^{th}$ step, the user who owns the credentials can selectively disclose attributes or statements about them in a publicly verifiable protocol. \par
The main objectives/design goals of this systems are to include threshold authorities, provide blind issuance and unlinkability, exhibit non-interactivity, liveness, and efficiency, using short credentials no matter the number of authorities or the number of attributes embedded in the credential.

\clearpage
\section{Notations and Assumptions}
\subsection{Zero Knowledge Proofs}
As described in Coconut\cite{sonnino_coconut:_2018}, the system uses non-interactive zero-knowledge proofs to assert knowledge and relations over discrete logarithm values. The notation used to represent them is the one introduced by Camenisch et al \cite{camenisch_proof_1997} as shown below:
\begin{equation}\nonumber
{\rm NIZK}\{(x,y,\dots): \textrm{statements about } x, y, \dots\}
\end{equation}

Note that all zero knowledge proof used are based on standard sigma protocols to show knowledge or representation of discrete algorithms. They are based on the DL assumption \cite{camenisch_proof_1997} and can be done without a trusted setup.

\subsection{Security Settings}
In addition, as detailed in Coconut\cite{sonnino_coconut:_2018}, the system requires groups $(\mathbb{G}_1,\mathbb{G}_2,\mathbb{G}_T)$ of prime order $p$ with a bilinear map $e:\mathbb{G}_1 \times \mathbb{G}_2 \rightarrow \mathbb{G}_T$ which satisfies the following properties: bilinearity, non-degeneracy, efficiency, and we use type 3 pairings which rely on the XDH (External Diffie-Hellman) Assumption, which in turn implies the difficulty of the co-CDH (Computational co-Diffie-Hellman) problem as well as the DDH (Decisional Diffie-Hellman) problem. In addition, DDH is only guaranteed in $\mathbb{G}_1$ and we would need to rely on the SXDH assumption (which is stronger) if we also want DDH in $\mathbb{G}_2$\cite{boneh_short_nodate}.

\clearpage
\section{Coconut System Design}
\subsection{Scheme Definitions}

The protocols that comprise the Coconut threshold credentials scheme are defined below as described in the Coconut paper\cite{sonnino_coconut:_2018}:

\begin{description}[leftmargin=1em, labelindent=0em]
\setlength\itemsep{0.5em}

\item[\definition{Setup($1^\lambda$)}{$params$}] 
defines the system parameters $params$ with respect to the security parameter $\lambda$. These parameters are publicly available.

\item[\definition{KeyGen($params$)}{$sk,vk$}] run by the authorities to generate their secret key $sk$ and verification key $vk$ from the public $params$.

\item[\definition{AggKey($vk_1, \dots, vk_t$)}{$vk$}] run by whoever wants to verify a credential to aggregate any subset of $t$ verification keys $vk_i$ into a single consolidated verification key $vk$. \textsf{AggKey} needs to be run only once.

\item[\definition{IssueCred($m,\phi$)}{$\sigma$}] Interactive protocol between a user and each authority, by which the user obtains a credential $\sigma$ embedding the private attribute $m$ satisfying the statement $\phi$. 

\item[\definition{AggCred($\sigma_1, \dots, \sigma_t$)}{$\sigma$}] run by the user to aggregate any subset of $t$ partial credentials $\sigma_i$ into a single consolidated credential.

\item[\definition{ProveCred($vk, m, \phi'$)}{$\Theta,\phi'$}] run by the user to compute a proof $\Theta$ of possession of a credential certifying that the private attribute $m$ satisfies the statement $\phi'$ (under the corresponding verification key $vk$).

\item[\definition{VerifyCred($vk, \Theta, \phi'$)}{$true/false$}] run by whoever wants to verify a credential embeding a private attribute satisfying the statement $\phi'$, using the verification key $vk$ and cryptographic material $\Theta$ generated by \textsf{ProveCred}. 
\end{description}

\clearpage
\subsection{Coconut Credential Scheme (specific to e-petition)}
This scheme is briefly described in the the \textit{applications} section of the Coconut paper\cite{sonnino_coconut:_2018} and can be seen as an extension of the work of Diaz et al \cite{diaz_privacy_2008} which does not consider threshold issuance of credentials. The below scheme is taken from the Coconut paper\cite{sonnino_coconut:_2018} however with slight modifications on the part of $\phi$ and $\phi'$ to make it specific to the e-petition.\par
It is important to clarify that the issuance of the credential happens only once, while the rest of the scheme can happen as many times as needed. The user must only randomize the credential and continue with the rest of the system scheme normally (basically starting at the \textsf{ProveCred} stage and continuing as usual every time the user wants to re-use the credential).

\begin{description}[leftmargin=1em, labelindent=0em]
\setlength\itemsep{0.5em}
\item[\definition{Setup($1^\lambda$)}{$params$}] Choose a bilinear group $(\mathbb{G}_1,\mathbb{G}_2,\mathbb{G}_T)$ with order $p$, where $p$ is a $\lambda$-bit prime number. Let $g_1, h_1$ be generators of $\mathbb{G}_1$, and $g_2$ a generator of $\mathbb{G}_2$. The system parameters are $params=(\mathbb{G}_1, \mathbb{G}_2, \mathbb{G}_T, p, g_1, g_2, h_1)$. \par

\textit{The \textsf{Setup} algorithm generates the public parameters. Credentials are elements of $\mathbb{G}_1$, while verification keys are elements of $\mathbb{G}_2$.}

\item[\definition{TTPKeyGen($params, t, n$)}{$sk,vk$}] Pick two polynomials $v,w$ of degree $t-1$ with coefficients in $\mathbb{F}_p$, and set $(x,y) = (v(0), w(0))$. Issue to each authority $i \in [1, \dots, n]$ a secret key $sk_i = (x_i,y_i) = (v(i), w(i))$, and publish their verification key $vk_i$ = $(g_2,\alpha_i,\beta_i) = (g_2,g_2^{x_i},g_2^{y_i})$. In addition, \textsf{TTPKeyGen} can be distributed using techniques of Gennaro~\etal~\cite{gennaro1999secure} or Kate~\etal~\cite{cryptoeprint:2012:377} (as in the Coconut paper\cite{sonnino_coconut:_2018}).

\item[\definition{IssueCred($m$)}{$\sigma$}] Credentials issuance is composed of three algorithms:
\begin{description}[leftmargin=1em, labelindent=0em]
\setlength\itemsep{0.5em}
\item \definition{PrepareBlindSign($m$)}{$d,\Lambda$} The users generate an \elgamal key-pair $(d, \gamma=g_1^{d})$; pick a random $o\in\mathbb{F}_p$,  compute the commitment $c_m$ and the group element $h\in\mathbb{G}_1$ as follows:
\begin{equation}\nonumber
c_m = g_1^m h_1^o \qquad{\rm and}\qquad h = \hashtopoint(c_m)
\end{equation} 
Pick a random $k \in \mathbb{F}_p$ and compute an \elgamal encryption of $m$ as below:
\begin{equation}\nonumber
c = Enc(h^m)=(g_1^k,\gamma^k h^m)
\end{equation}
Output $(d, \Lambda=(\gamma, c_m, c, \pi_{s}))$, where $\pi_{s}$ is defined by:
\begin{eqnarray}\nonumber
\pi_{s} &=& {\rm NIZK}\{(d, m, o, k): \gamma = g_1^d \;\land\; c_m=g_1^mh_1^o\\ \nonumber
 && \land\; c = (g_1^k,\gamma^k h^m) \;\land\;  \phi(m)=1\}
 \end{eqnarray}

\textit{To keep an attribute $m \in \mathbb{F}_p$ hidden from the authorities, the users use a blinding factor $o$ to keep it blind from adversaries who would try expected values of m. Then the users encrypt it with their El-Gamal secret key so that the authorities are able to blindly sign the credential. They must also include a zero-knowledge proof ensuring correctness of $\gamma, c_m,c$}

\item \definition{BlindSign($sk_i, \Lambda, \phi$)}{$\tilde{\sigma}_i$} The authority $i$ parses $\Lambda=(\gamma, c_m, c, \pi_{s})$, $sk_i=(x,y)$, and $c=(a,b)$. Recompute $h = \hashtopoint(c_m)$. Verify the proof  $\pi_{s}$ using $\gamma$, $c_m$ and $\phi$; if the proof is valid, build $\tilde{c}=(a^y,h^xb^y)$ and output $\tilde{\sigma}_i = (h, \tilde{c})$; otherwise output $\perp$ and stop the protocol.

\textit{To blindly sign the attribute, each authority $i$ verifies the proof $\pi_{s}$, and uses the homomorphic properties of El-Gamal to generate an encryption $\tilde{c}$ of $h^{x_i+y_i\cdot m}$ as below:}
\begin{equation}\nonumber
\tilde{c} = (a^y, h^{x_i} b^{y_i}) = (g_1^{ky_i}, \gamma^{ky_i}h^{x_i+y_i\cdot m})
\end{equation}
\textit{Note that every authority must operate on the same element $h$. Intuitively, generating $h$ from $h=\hashtopoint(c_m)$ is equivalent to computing $h=g_1^{\tilde{r}}$  where $\tilde{r} \in \mathbb{F}_p$ is unknown by the users (as in Pointcheval and Sanders) 
And since $h$ is deterministic, every party is able to derive it in isolation.}

\item \definition{Unblind($\tilde{\sigma}_i, d$)}{$\sigma_i$} The users parse $\tilde{\sigma}_i=(h, \tilde{c})$ and $\tilde{c}=(\tilde{a},\tilde{b})$; compute $\sigma_i = (h,\tilde{b}(\tilde{a})^{-d})$. Output $\sigma_i$.
 \end{description}

\textit{Upon reception of $\tilde{c}$, the users decrypt it using their El-Gamal private key $d$ to recover the partial credentials $\sigma_i = (h, h^{x_i+y_i\cdot m})$}
 
\item[\definition{AggCred($\sigma_1, \dots, \sigma_t$)}{$\sigma$}] Parse each $\sigma_i$ as $(h,s_i)$ for $i \in [1, \dots, t]$. Output $(h,\prod^t_{i=1} s_i^{l_i})$, where $l$ is the Lagrange coefficient:
\begin{equation}\nonumber
l_i = \left[\prod^t_{i=1, j\neq i} (0-j)\right] \left[\prod^t_{i=1,s j\neq i} (i-j)\right]^{-1} \;{\rm mod}\; p
\end{equation}

\textit{The users then aggregate any subset of $t$ partial credentials. This algorithm uses the Lagrange basis polynomial $l$ which allows to reconstruct the original $v(0)$ and $w(0)$ through polynomial interpolation;}

\begin{equation}\nonumber
v(0) = \sum^t_{i=1} v(i)l_i \quad {\rm and} \quad w(0) = \sum^t_{i=1} w(i)l_i
\end{equation}

\textit{
However, this computation happens in the exponent---neither the authorities nor the users should know the values $v(0)$ and $w(0)$. One can easily verify the correctness of \textsf{AggCred} of $t$ partial credentials $\sigma_i=(h_i,s_i)$ as below.}

\begin{eqnarray} \nonumber
	s &=& \prod^t_{i=1} \left(s_i\right)^{l_i} = \prod^t_{i=1} \left(h^{x_i+y_i\cdot m}\right)^{l_i} \\ \nonumber
	&=& \prod^t_{i=1} \left(h^{x_i}\right)^{l_i} \prod^t_{i=1} \left(h^{y_i\cdot m}\right)^{l_i} = \prod^t_{i=1} h^{(x_i l_i)} \prod^t_{i=1} h^{(y_i l_i)\cdot m} \\ \nonumber
	&=& h^{v(0)+w(0)\cdot m} = h^{x+y\cdot m}
\end{eqnarray}

\item[\definition{ProveCred($vk, m, \sigma, \zeta$)}{$\Theta, \zeta$}] Parse $\sigma=(h,s)$ and $vk=(g_2,\alpha,\beta)$. Pick at random $r',r \in \mathbb{F}_p^2$; set $\sigma'=(h',s')=(h^{r'},s^{r'})$; build $\kappa = \alpha\beta^m g_2^r$, $\nu=\left(h'\right)^r$ and $\zeta=g_s^{m}$, where this identifier is generated through a hash function on the  petition identifier $\mathbb{F}_p \rightarrow \mathbb{G}_1:\widetilde{H}(s)=g_s \;|\; s\in\mathbb{F}_p$. Output $(\Theta=(\kappa, \nu, \sigma',\pi_v),\zeta)$, such that $\pi_v$ is:
\begin{equation}\nonumber
    \pi_v={\rm NIZK}\{(m,r): \kappa=\alpha\beta^m g_2^r \ \land \ \nu=\left(h'\right)^r \ \land \  \zeta=g_s^{m}\} 
\end{equation}

\textit{Before verification, the verifier collects and aggregates the verifications keys of the authorities---this process  happens only once and ahead of time. First, the users randomize the credentials by picking a random $r' \in \mathbb{F}_p$ and computing $\sigma'=(h',s')=(h^{r'},s^{r'})$; then, they compute $\kappa$ and $\nu$ from the attribute $m$, a blinding factor $r\in\mathbb{F}_p$ and the aggregated verification key:}
\begin{equation*}
\kappa=\alpha\beta^m g_2^r \qquad{\rm and}\qquad \nu=(h')^r
\end{equation*}
\textit{Finally, they send $\Theta=(\kappa, \nu, \sigma', \pi_v)$ and $\zeta$ to the verifier where $\pi_v$ is a zero-knowledge proof asserting the correctness of $\kappa$  and $\nu$; and that the private attribute $m$ embedded into $\sigma$ satisfies $\zeta=g_s^{m}$. The proof $\pi_v$ also ensures that the users actually know $m$ and that $\kappa$ has been built using the correct verification keys and blinding factors.}

\item[\definition{VerifyCred($vk, \Theta, \zeta$)}{$true/false$}] Parse $\Theta = (\kappa, \nu, \sigma',\pi_v)$ and $\sigma'=(h',s')$; verify $\pi_v$ using $vk$ and $\zeta$. Output $true$ if the proof verifies, $h'\neq1$ and $e(h',\kappa)=e(s'\nu,g_2)$; otherwise output $false$.
\end{description}

\textit{The pairing verification is similar to Pointcheval and Sanders \cite{pointcheval_short_2015} and Boneh et al\cite{boneh_short_nodate}; expressing $h'=g_1^{\tilde{r}} \; | \; \tilde{r} \in \mathbb{F}_p$, the left-hand side of the pairing verification can be expanded as:}
\begin{equation*}
    e(h',\kappa) = e(h',g_2^{(x+my+r)}) = e(g_1,g_2)^{(x+my+r) \tilde{r}}
\end{equation*}
\textit{and the right-hand side:}
\begin{equation*}
    e(s' \nu,g_2) = e(h'^{(x+my+r)},g_2) = e(g_1,g_2)^{(x+my+r)  \tilde{r}}
\end{equation*}
\textit{From where the correctness of \textsf{VerifyCred} follows.}

\vspace{20mm}

The main difference between this specific e-petition Coconut credential scheme and the generic Coconut credential scheme described in the Coconut paper\cite{sonnino_coconut:_2018}, which can be found in the appendix \nameref{coconut_gen} section, is that the $\phi$ parameter in \textsf{PrepareBlindSign} is simply equal to 1 and can be ignored because it's not needed and $\phi'(m)=1$ parameter in \textsf{ProveCred} and \textsf{VerifyCred} is implemented as $\zeta=g_s^m=H^*(s)^m$ which is specific to the e-petition and basically prevents users from voting multiple times (also known as the double spend problem). Everything else is essentially the same.

\vspace{20mm}

\subsection{Signature Petition Scheme}
The signature petition scheme implemented here is one consisting of an additive homomorphic encryption system along with a threshold decryption technique on the encrypted result.\par

Once the the user's credentials are verified, the user will be elligible to participate in the petition voting. To optimize, it is done in parallel with the \textsf{ProveCred} and \textsf{VerifyCred} protocols above. The vote output is sent along with the \textsf{ProveCred} output. At the authority's side, if the \textsf{VerifyCred} is valid (credential valid and there is no double spend attempt), then they continue with the rest of the vote procedure.

In this system, there are 3 types of parties, the voters/users, the decrypting authorities, and the petition entity (owner/creator). Note that these decrypting authorities are not necessarily the same as the signing authorities in the previous Coconut procedure. Technically for simplicity, authorities can perform both signing and decrypting functions but they do not neccessarily have to be merged at all. Firstly, the authorities generate their El-Gamal keypairs ($d$, $\gamma=g^d$) = (secret key: $sk$, verification key: $vk$), and publish their respective public verification keys. \par

The users then collect and aggregate the El-Gamal public keys of the the authorities. For simplicity here, we use an n-out-of-n credentials system with regards to the multiple authorities.

$$\gamma_{agg} =(\prod_{i=0}^{N-1} \gamma_i)$$

In addition to this, the authorities must provide a NIZK proof of the secret corresponding to the public key. This is done mainly to prevent a malicious \textit{rogue public key attack} \cite{noauthor_bls_nodate} whereby the malicious actor $eve$ broadcasts a \textit{fake} public key they when the aggregated with the rest, produces a key that is actually the real public key of the malicious actor, as shown below:

$$\gamma_{fake} = \gamma_{eve} \cdot (\prod_{i=1}^{N-1} \gamma_i)^{-1}$$

Then, when aggregated with the rest of the public keys, the aggregated key will produce the following:
$$\gamma_{agg} = (\prod_{i=0}^{N-1} \gamma_i) = \gamma_{fake} \cdot (\prod_{i=1}^{N-1} \gamma_i) = \gamma_{eve} \cdot (\prod_{i=1}^{N-1} \gamma_i)^{-1} \cdot (\prod_{i=1}^{N-1} \gamma_i) = \gamma_{eve}$$ 

The consequences of this attack are that the group's public key would then be in control of the attacker $Eve$. Authorities then provide a NIZK proof of knowledge of the secret key to the public key they publish to prevent this attack from occurring:

$$\pi_\gamma = \text{NIZK}\{ (d): \gamma=g^d \}$$

Once the users have the aggregated verification key, they El-Gamal encrypt their vote $v$ with the aggregated verification key.

$$\text{choose random } k \in \mathbb{F}_p$$
$$enc(v_i)=(g^k, \gamma_{agg}^k \cdot h^{v_i})=(a, b)$$

An important thing to note is that since the vote is encrypted and only viewable by the voter themselves, a problem arises being that a mallicous voter can encrypt an arbitrary vote to adversarily affect the outcome. To circumvent this, a zero-knowledge proof $\pi_p$ is added to ensure that the vote $v$ is either $0$ or $1$, namely $v(v-1)=0$, such that:

$$\pi_p = \text{NIZK}\{ (v, k): c = (g^k, \ \gamma_{agg}^k \cdot h^v) \ \land \ v(1-v)=0 \}$$
\begin{center}
{\scriptsize (Exactly how this is done is detailed in the \nameref{impl} section below)}
\end{center}

Note the following two points. First, this NIZK proof does not prevent \textit{blank} votes (ie. 0-0 or 1-1). Second, that we also include the encryption of the vote inverse for clarity/simplicity reasons, so that we get a total of $1$ votes and $0$ votes when we decrypt even though the inverse of the vote can be derived from the vote as done with the Coconut implementation in Python\cite{coconut_python}. Then the users send that to the petition entity
$$enc_{not}(v_i)=(a^{-1}, b^{-1}\cdot h)=(g^{-k}, \gamma_{agg}^{-k} \cdot h^{-v_i} \cdot h)=(g^{-k}, \ \gamma_{agg}^{-k} \cdot h^{1-v_i})$$

The user then sends all of this information, $(enc(v_i), enc_not(v_i), \pi_p)$ to the petition entity. \par

The petition entity then homomorphically adds the encrypted votes by simply multiplying them together, $N$ being the total number of voters. 
$$enc(total(v))=\prod_{i=0}^{N-1}enc(v_i)=(\prod_{i=0}^{N-1}a, \prod_{i=0}^{N-1}b) = (\prod_{i=0}^{N-1}g^{k_i}, \prod_{i=0}^{N-1}(\gamma_{agg}^{k_i} \cdot h^{v_i}))$$
$$= (g^{\sum_{i=0}^{N-1}k_i}, \gamma_{agg}^{\sum_{i=0}^{N-1}k_i} \cdot h^{\sum_{i=0}^{N-1}v_i}) = (g^{k_{agg}}, \gamma_{agg}^{k_{agg}} \cdot h^{\sum_{i=0}^{N-1}v_i})=(\tilde{a}, \ \tilde{b})$$

Similarly, for $enc_{not}(v_i)$ we get:
$$enc(total_{not}(v))=\prod_{i=0}^{N-1}enc_{not}(v_i)=(g^{k_{agg}}, \gamma_{agg}^{k_{agg}} \cdot h^{\sum_{i=0}^{N-1}(1-v_i)})=(\tilde{a}_{not}, \ \tilde{b}_{not})$$

After the petition has ended, to get the actual results from the total encyrpted result, the petition entity must send this total to the decrypting authorities who jointly decrypt the total with their secret key one by one:

$$total(v) = (\tilde{a}, \ \tilde{b} \cdot \tilde{a}^{-d_i})$$
\begin{center}
{\scriptsize (And similarly for $total_{not}(v)$)} 
\end{center}

Once the last decrypting authority has jointly decrypted, the result will look like this:

$$total(v) = (\tilde{a}, \ h^{\sum_{i=0}^{N-1}v_i})$$
$$total_{not}(v) = (\tilde{a}, \ h^{\sum_{i=0}^{N-1}(1-v_i)})$$

Then all the last decrypting authority has do to get  actual result would be to calculate the discrete logarithm of $h^{total}$. Discrete logarithm computations are usually very tough to do, however in this case it should not be too hard since we know that all values must be either $0$ or $1$ and we can compute $h^i$ for possible outcome values of $i$ until $h^i \ = \ h^{total}$. It works because in this case, $total$ is a small enough value for this to be done.

\clearpage
\section{System Architecture and Implementation}
\subsection{Starting Points}
\subsubsection{Fork of JavaScript E-Cash Tangerine Implementation}
For the implementation  of this project, the starting point was to fork the repository\cite{stuczynski_multi-authority-sdc:_2018} of a previous fellow UCL master's student, Jedrzej Stuczynski, who implemented a multi authority selective disclosure credential system for an e-cash scheme. His implementation was based on the Tangerine Signature System, which exhibited some shortcomings in terms of security and privacy when compared to Coconut, discussed in greater detail in the \nameref{tang} section. Accordingly, his implementations included these flaws/limitations, and are addressed below.
\par
The first step to beginning this project from the forked repository, was to remove unneeded aspects/components such as the extra coin credentials (value, TTL, ID). The next step was to fix these issues and limitations. Once I had done that, I could continue from that starting point to implement the rest of the system onward from the \textsf{IssueCred} step, since the rest of the design/implementation is different.
\par

\subsubsection{Criticism of Previous Implementation}
Even though Jedrzej's implementation was on the most part very properly done, it still exhibited some flaws that are worth nothing but most importantly had to be dealt with.\par

First of all, the commitment to the secret credential did not include any blinding factor as mentioned above in the \nameref{tang} section, which is detrimental to the security of the system. In addition, the implementation of the commitment was done in the group $\mathbb{G}_2$ using the generator $g_2$. This was changed to be done in the group $\mathbb{G}_1$ using the generator $g_1$. The system still worked with the computations/constructions in $\mathbb{G}_2$, however they incurred an unneeded extra cost/overhead. This extra cost/overhead includes the fact that the commitment, credentials, and encryptions are unnecessarily twice as large. Moreover, arithmetic operations (ie. point addition and scalar multiplication) are more expensive (twice as much) in $\mathbb{G}_2$. This is illustrated more in the \nameref{testing} section.\par

Furthermore, the implementation did not include a zero-knowledge proof for the El-Gamal secret key of the client, the random $k$ used in this encryption, and for the blinding factor $o$, obviously because it was not even in the system design to begin with. \par

\subsection{Scheme Architecture}
The architecture consists of 3 different types of parties, the client/user, the signing authorities, and the petition owner/creator(s), as can be seen in figure \ref{fig:petition} below. Note that the ledger is acts as a layer in between citizen/user and petition creator/owner communications. \par

\begin{figure}[h]
    \centering
    \includegraphics[width=.6\textwidth]{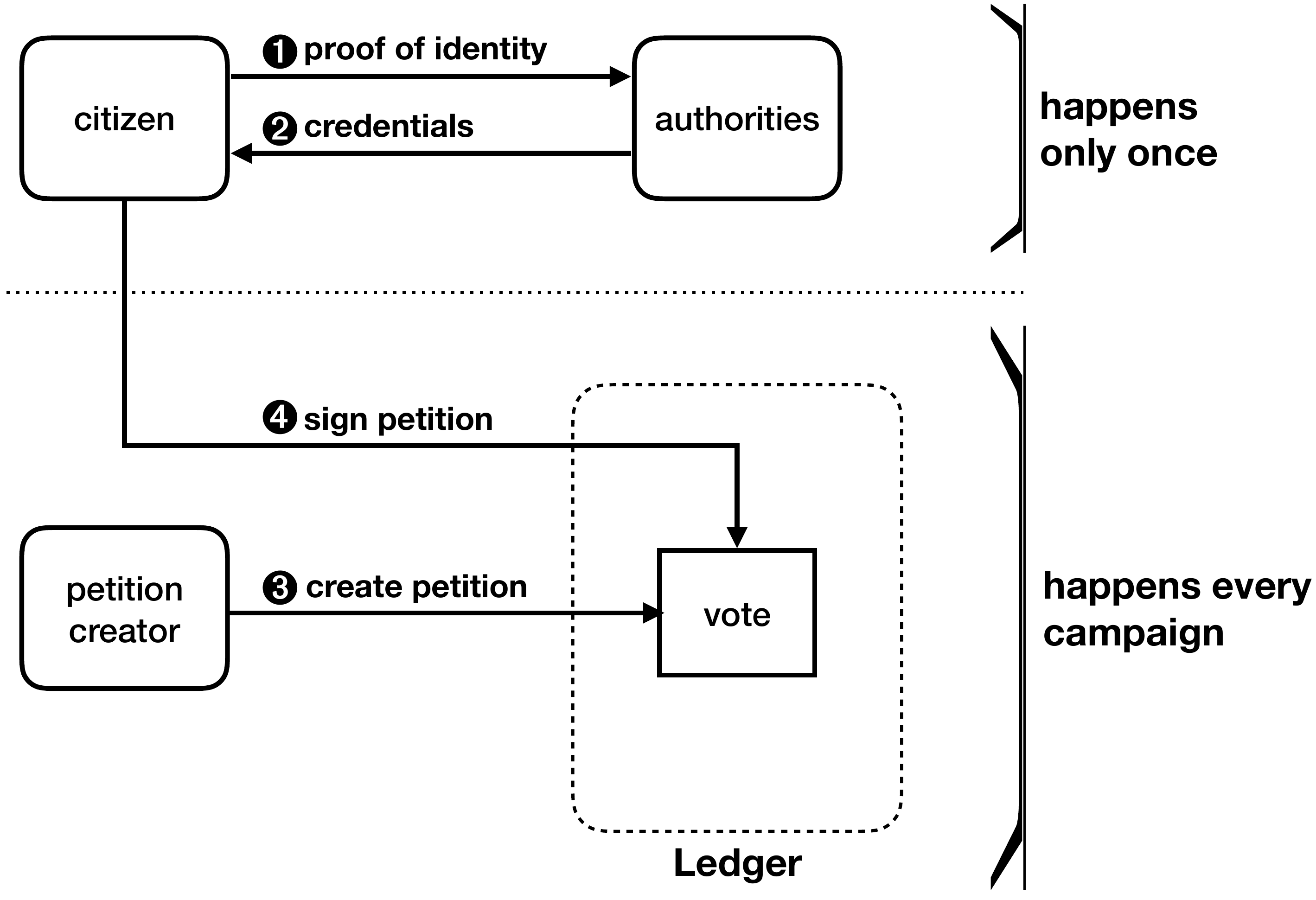}
    \caption{Coconut Petition Diagram\cite{sonnino_coconut:_2018}}
    \label{fig:petition}
\end{figure}

\subsubsection{Platforms/Languages}
The main motivation behind using JavaScript for the system was so that the client is able to locally and trustlessly compute the necessary cryptographic constructs and operations without relying on any 3rd party to do them on their behalf. As well as to be able to have an appealing graphical user interface. To this end, JavaScript was the most suitable choice.

The client is implemented through a React web application (JavaScript) and the rest of the entities are implemented through Node.js servers simulations which can be replaced by something more sophisticated such as a blockchain platform like ChainSpace (discussed further in \nameref{conc}).
\par

All components are written in the JavaScript implementation of the ECMAScript 2017, or ES8, standard in order to easily write, share, and reuse modules between the different entities of the system. Then at run-time transpiled to ECMAScript 5, or ES5, in order to accommodate for browser vendors which are slow to adopt these new language features.\par

\subsubsection{Cryptography}
Regarding the cryptography of the system, the Milagro Crypto JavaScript library\cite{scott_apache_nodate} is used for both the client and the servers. Specific components used include the SHA256 hashing algorithm, finite field operations (BIG number types), elliptic curve point types (ECP and ECP2 types) and their corresponding operations/computations. The setting used is the Fp254BNb and Fp254n2BNb 256-bit Barreto-Naehrig curves \cite{barreto_pairing-friendly_2006} that the authors of the library implemented using methods presented by Aranha et al\cite{aranha_faster_2011}. Last but not least, Milagro's implementation uses the Marsaglia and Zaman random number generator\cite{marsaglia_new_1991},
and it's security relies on the being seeded with an external source of entropy. In this case, we also use JavaScript's WebCryptoAPI\cite{noauthor_web_nodate}.
\par

\subsubsection{Communication}
The system entities/parties communicate with each other through the Representational State Transfer (REST) architecture using GET and POST HTTP requests and responses. One thing to note is that since the cryptographic objects exchanged between parties are transformed to an array of bytes when sent and then recomposed by the reciever. This is because, as well as requiring less bandwidth, the cryptographic objects are represented using the Milagro\cite{scott_apache_nodate}
library types, which have lengthy chains of dependencies, and would cause issues when attempting to call methods on them due to the effects of serialization and deserialization.

\subsection{Scheme Implementation} \label{impl}
The code has been released as open source software on GitHub\cite{jadwahab_coconut-petition:_2018} under the Apache 2.0 license.

\subsubsection{Setup}
When the setup is initialized, all parties generate their corresponding cryptographic keys/keypairs along the proper parameters defined above. 
Note that since the scheme we implement is and n-out-of-n (as opposed to threshold) for the sake of simplicity, we do not need the \textsf{TTPKeyGen} step. 
Then, corresponding parties query the rest of the parties for their respective public keys. An example of the client generating their El-Gamal keypair, as well as their device keypair (shown in an array of bytes) and querying the rest of the parties for their public keys is shown in figure \ref{fig:setup} below:

\begin{figure}[h!]
    \centering
    \includegraphics[width=1\textwidth]{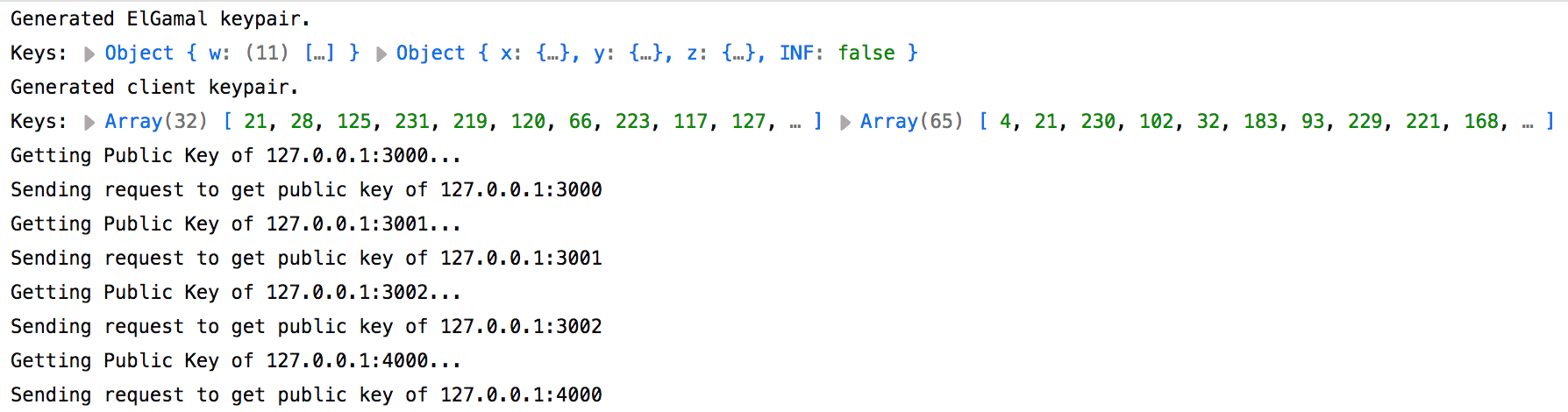}
    \caption{Client setup}
    \label{fig:setup}
\end{figure}

\subsubsection{IssueCred}
PrepareBlindSign:\par
When clients click on the "Sign Credential" button shown below, 
the \textsf{PrepareBlindSign} algorithm runs as described in the system design and the output $\Lambda$ is sent to each of the authorities.
However the only added procedure is that the client's device public key is added to the input of the hash along with the commitment and is included in the zero-knowledge proof for obvious security reasons.

    \begin{figure}[h!]
        \centering
        \includegraphics[width=.7\textwidth]{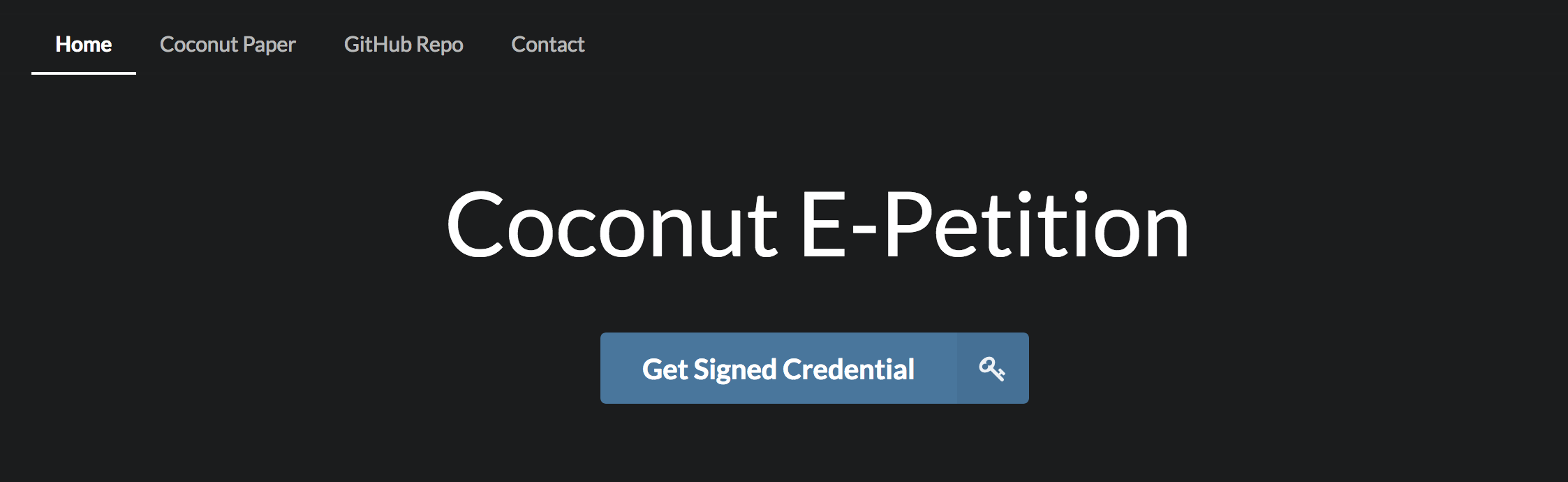}
        \caption{Home page}
    \end{figure}

As shown below in figure \ref{fig:cred-obj} JSON object of 5 elements is sent: \textit{pk\_cred\_bytes}, \textit{pk\_client\_bytes}, \textit{enc\_sk\_bytes}, \textit{requestSig}, and \textit{proof}. As explained above, components being sent are first transformed into an array of bytes. The \textit{}{pk\_cred\_bytes} is the commitment (or public key) of the secret credential. The \textit{pk\_client\_bytes} is the device public key of the client. The \textit{enc\_sk\_bytes} is the El-Gamal encryption of the secret credential. The \textit{requestSig} is the ECDSA signature of the client with the client's device secret key. Finally, the \textit{proof} is the output of the zero-knowledge proof $\pi_s$.
    
    \begin{figure}[h!]
        \centering
        \includegraphics[width=.9\textwidth]{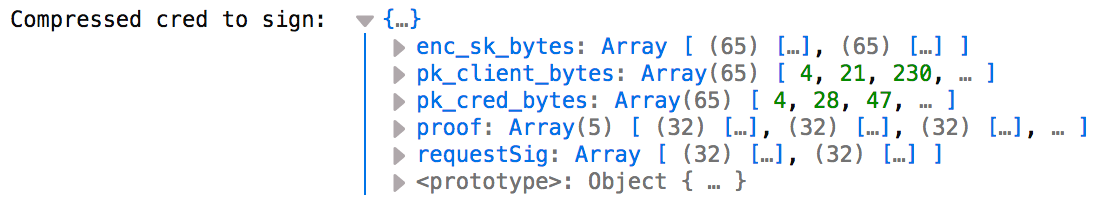}
        \caption{Credential object to be signed by authorities}
        \label{fig:cred-obj}
    \end{figure}

\noindent BlindSign:\par
When a signing authority gets the POST request from the client, they first check the validity of the signature. Then they check the validity of the zero-knowledge proof. If both of these conditions hold, then they sign the credential with their secret keys and send the partial encrypted credential back to the client.

    \begin{figure}[h!]
        \centering
        \includegraphics[width=.7\textwidth]{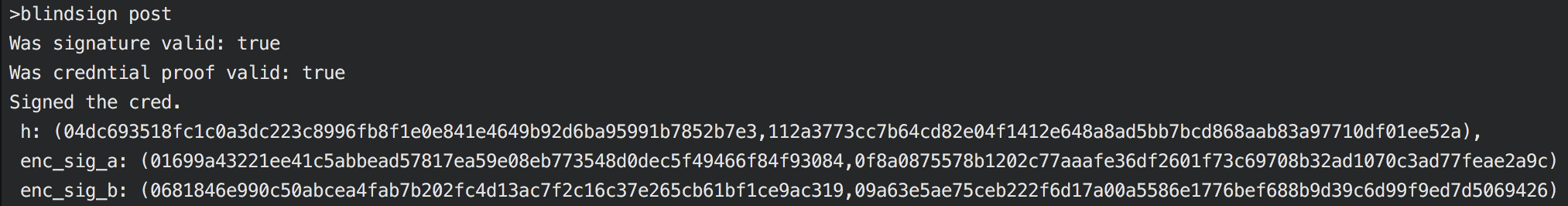}
        \caption{Blind signing of credential}
        \label{fig:blindsign}
    \end{figure}

\noindent Unblind: \par
When the client gets the response from the signing authority, they decrypt it with their El-Gamal secret key.

\subsubsection{AggCred}
Following that, the aggregation of the partial signatures is done by the client in the browser.

\subsubsection{ProveCred}
Once the above steps are completed, the "Sign Credential" button turns into a "Randomize Credential" button as shown below in figure \ref{fig:rand-cred}.

    \begin{figure}[h!]
        \centering
        \includegraphics[width=.7\textwidth]{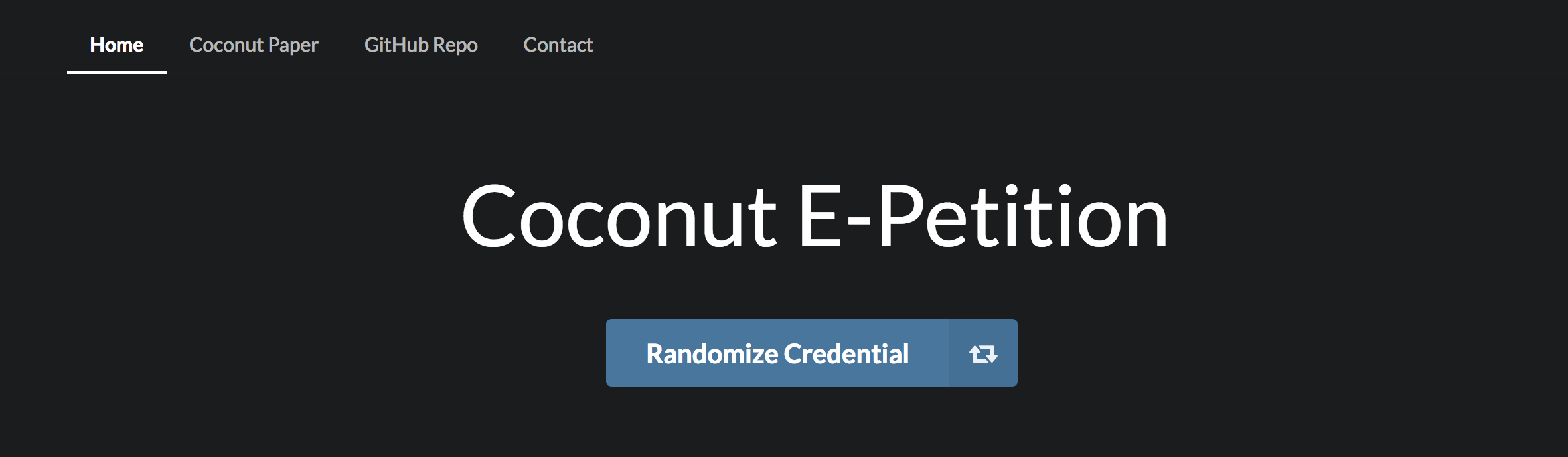}
        \caption{Client randomize credential}
        \label{fig:rand-cred}
    \end{figure}

When the button is clicked, it becomes disabled and an input form is added to the DOM (Document Object Model) which lets the client vote "yes/for" or "no/against" for a specific petition by clicking either the thumbs up or the thumbs down buttons respectively. This is shown below in figure \ref{fig:input-form} They must also input the specific ID of the petition they would like to vote on. Note that this input petition ID is what is hashed and raised to power of the secret credential in order to create $\zeta$, which prevents users from voting for the same petition more than once. 

    \begin{figure}[h!]
        \centering
        \includegraphics[width=.7\textwidth]{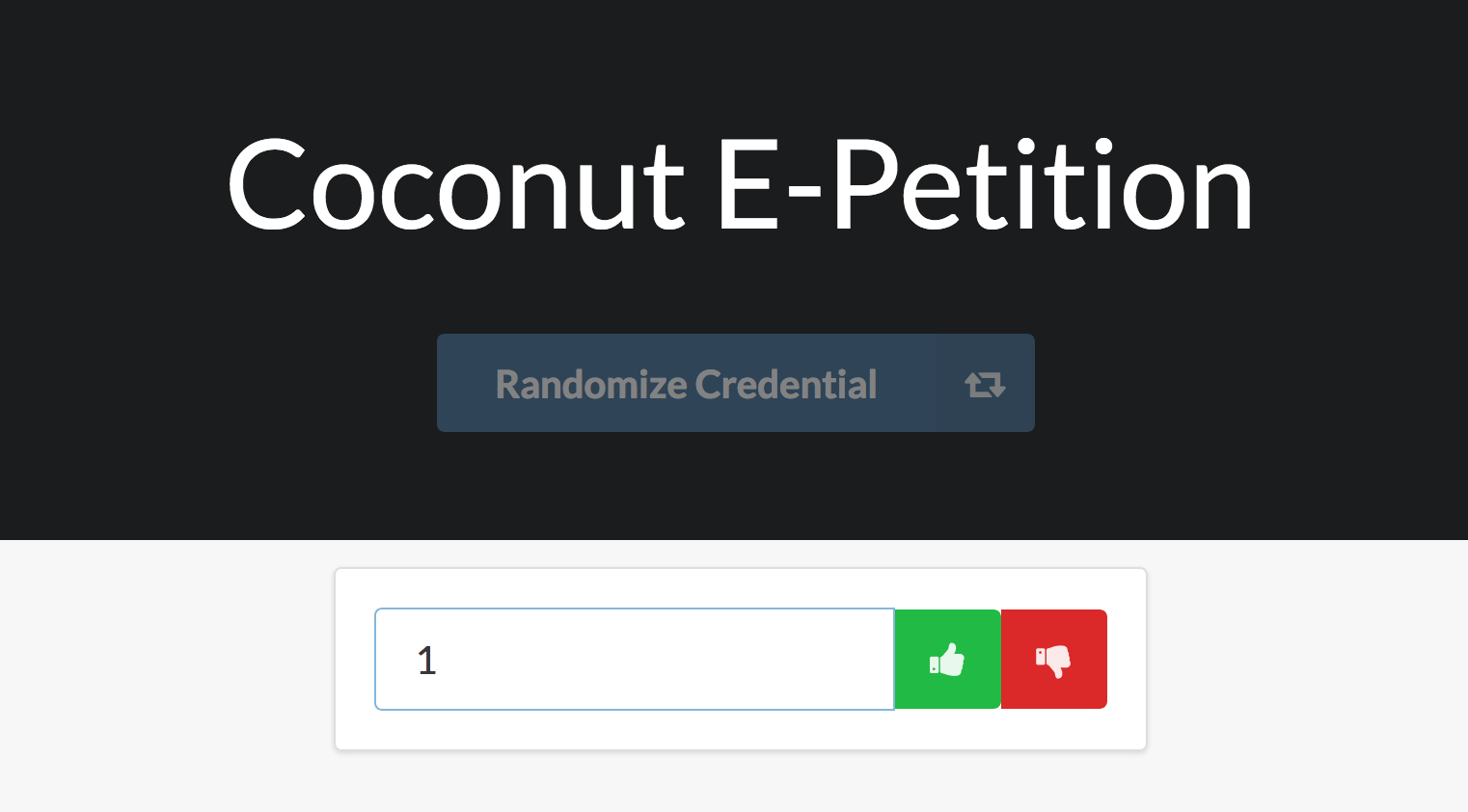}
        \caption{Client inputs the \textit{petition ID} and votes "yes" or "no"}
        \label{fig:input-form}
    \end{figure}

\begin{figure}[h!]
        \centering
        \includegraphics[width=.7\textwidth]{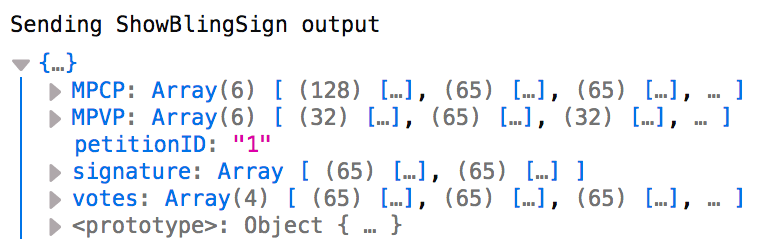}
        \caption{\textsf{ProveCred} and \textsf{Vote} scheme outputs}
        \label{fig:prove-cred}
    \end{figure}

As shown in figure \ref{fig:prove-cred}, a JSON object of 5 elements is sent: \textit{MPCP}, \textit{MPVP}, \textit{peitionID}, \textit{signature}, and \textit{votes}. As explained above components being sent are first transformed into an array of bytes. The \textit{MPCP} (make-proof-credential-petition) is the output of the zero-knowledge proof $\pi_v$. The \textit{MPVP} (make-proof-vote-petition) is the output of the zero-knowledge proof $\pi_p$. The \textit{peitionID} is the Id of the specific petition. The \textit{signature} is the client's randomized signature. Finally, the \textit{votes} element is the combination of the encrypted vote and vote inverse.

When getting a success response from the petition owner/creator entity, the button will show which petition you voted for, thumbs up or down, and the "Randomize Credential" button will turn back on as shown in figure \ref{fig:vote} below.

    \begin{figure}[h!]
        \centering
        \includegraphics[width=.7\textwidth]{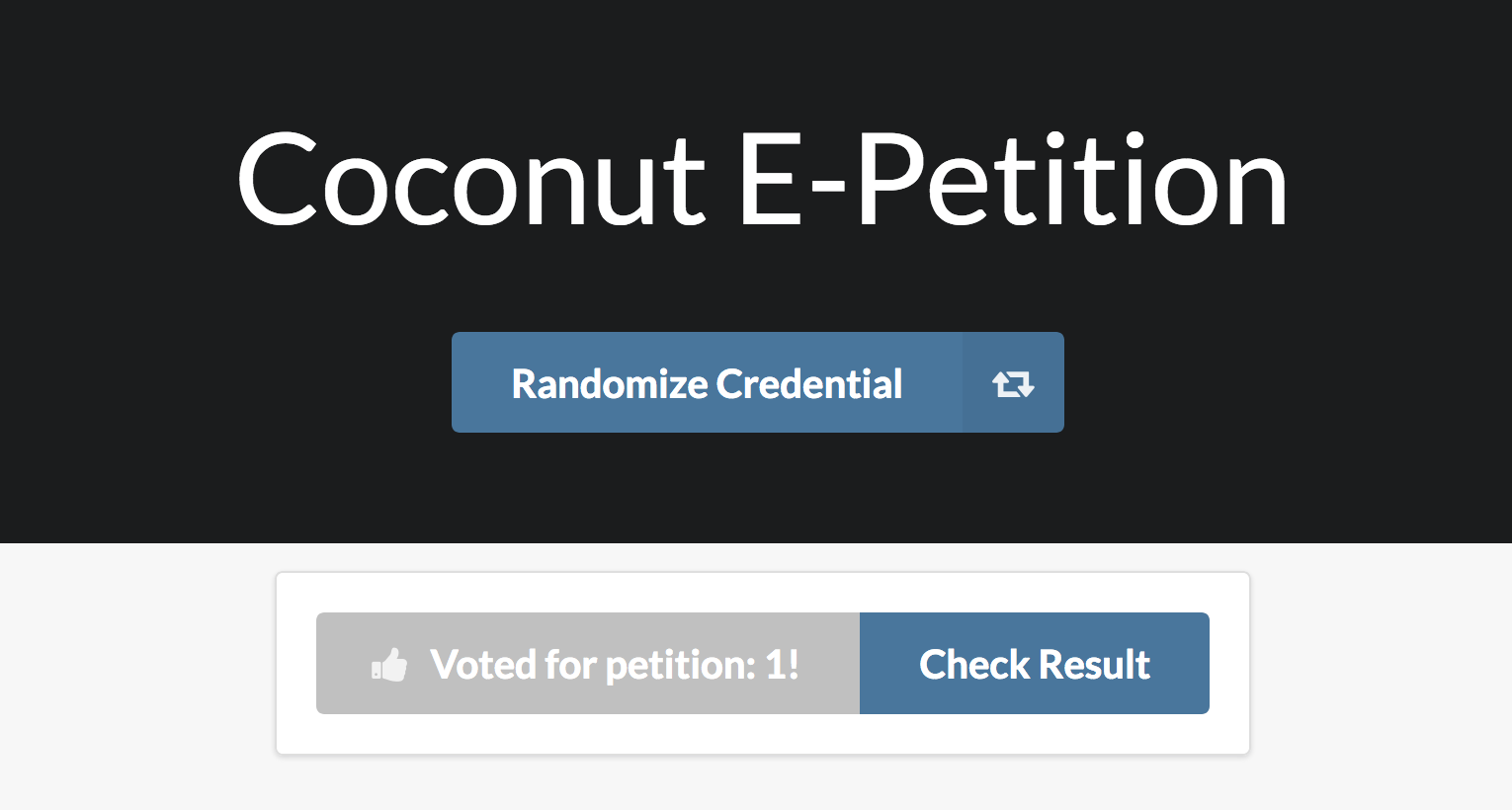}
        \caption{Client vote}
        \label{fig:vote}
    \end{figure}

If the client clicks on the "Check Result" button, the client will query the petition owner for the result of said petition. In the case of the screenshot below, the petition owner replied that the petition had not yet ended. Also, if the client tried to vote for the same petition, the petition owner will know from the repeated value of $\zeta$ and prevent them from voting for the same petition. This is shown in figure \ref{fig:vote-same} below. Note that if the client tries changing the value of $\zeta$ then the proof of knowledge (MPCP) will become invalid and also prevent them from voting.

    \begin{figure}[h!]
        \centering
        \includegraphics[width=.7\textwidth]{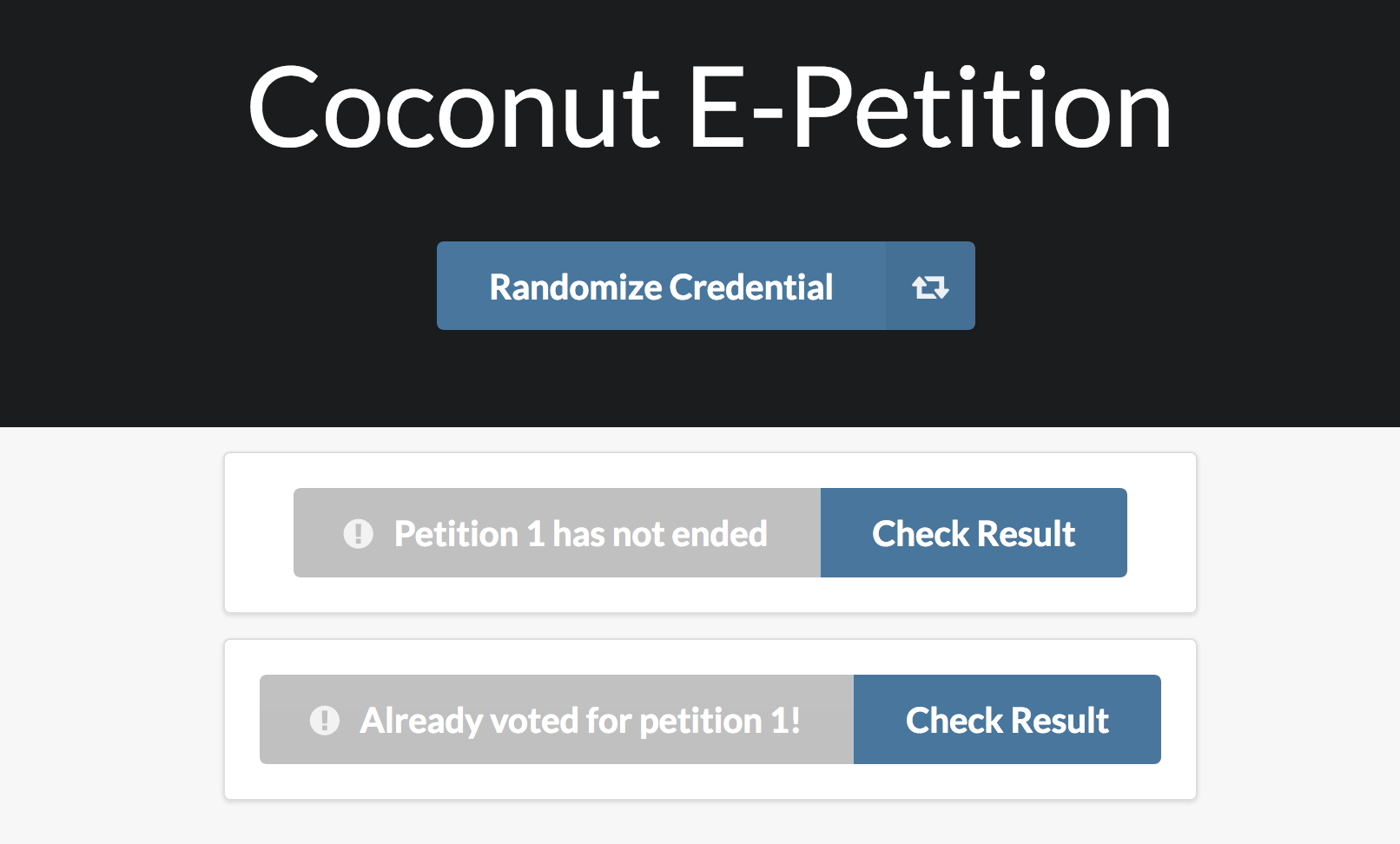}
        \caption{Client check result \& vote for same petition}
        \label{fig:vote-same}
    \end{figure}

Once the petition ends, the petition owner adds all of the votes together homomorphically and then sends it to one of the decrypting authorities. The authorities will decrypt the total with their El-Gamal secret key and forward it to the next decrypting authority until the last authority is reached. Once the total reaches the final authority, they decrypt it and retrieve the log to get the actual values of the total votes ("yes/for") and total vote inverses ("no/against"). Finally they send that back to the petition owner who will get the final results of the specific petition to give to the clients when queried for the result. The client then simply displays the result of the petition in the form "yes" total -"no" total. The result is displayed in green if the petition passes, and in red if the petition fails as shown in figure \ref{fig:result} below.

    \begin{figure}[h!]
        \centering
        \includegraphics[width=.7\textwidth]{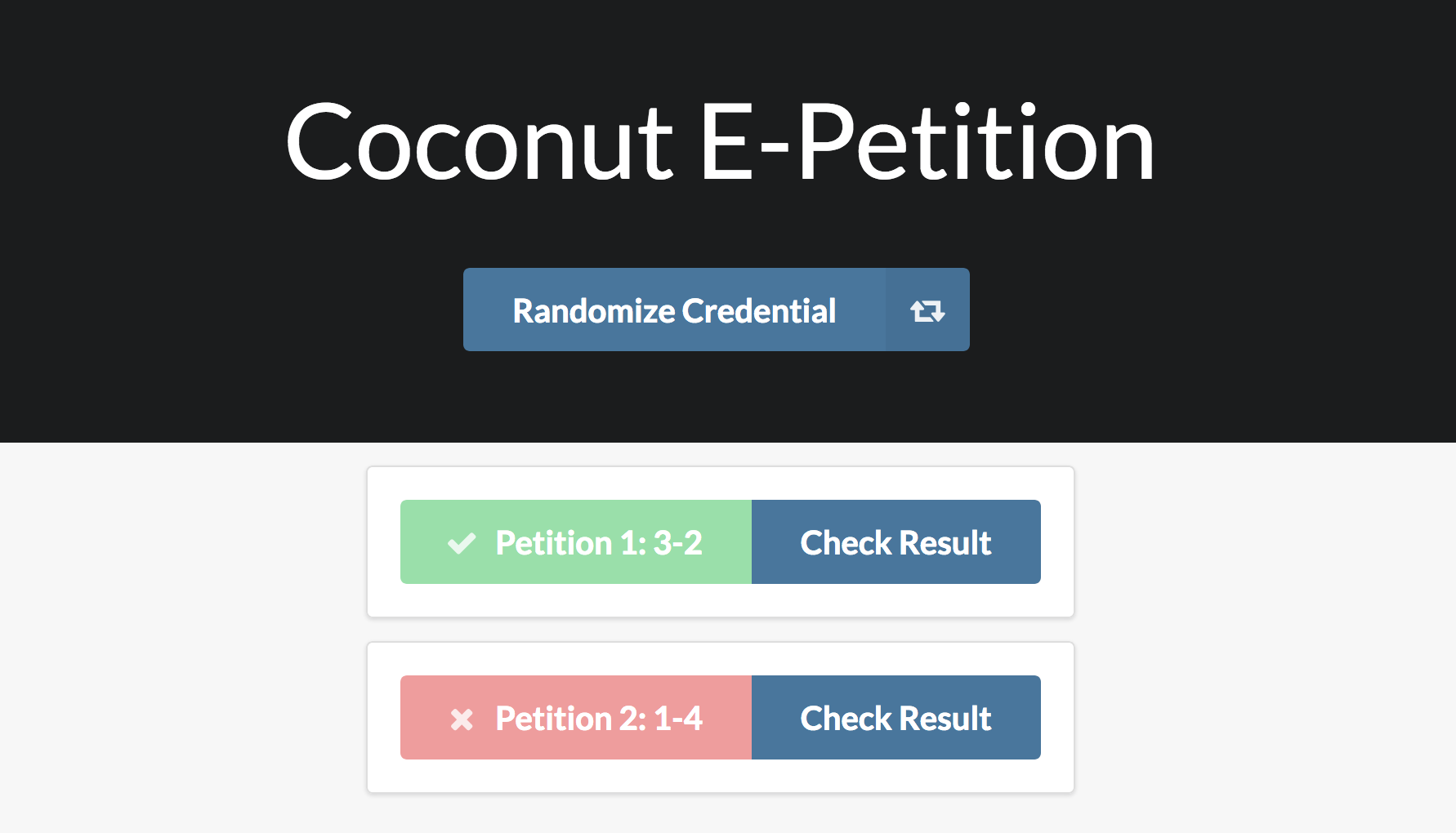}
        \caption{Client check result after petition ends}
        \label{fig:result}
    \end{figure}

\textbf{A very important thing to note} here is that this process described is sub-optimal in terms of security and verifiability since the clients would have to trust the signing authorities and the petition owner. It was only done this way for this prototype implementation for the sake of simplicity. An improved alternative would have for example the client get the total encrypted votes and then query each decrypting authority one by one themselves and do the log computation themselves. Another more elegant alternative would be to use a public blockchain for the results which would provide transparency and total public verifiability, such as the case with the Ethereum implementation of the Open Vote Network\cite{kiayias_smart_2017}. We discuss this in further detail in the final section of \nameref{conc}.

\clearpage
\section{Testing and Evaluation}\label{testing}
The testing of the system was focused on breaking down the system into smaller parts for unit testing of specific functions of the system. The unit tests relied on the \textit{Mocha}\cite{noauthor_mochajs/mocha:_nodate} JavaScript test framework as well as the \textit{Chai}\cite{noauthor_chaijs/chai:_nodate} assertion framework that can be paired with any testing framework. The unit tests were done to guarantee correctness of mainly the cryptographic constructions and operations without the overhead of the rest of the system and communications. Note that testing for the client and the servers was the same as they both relied on the same modules for the cryptographic constructions and operations.\par

Performance was measured according to some of the tests created for  functionality. As in the Tangerine system\cite{stuczynski_tangerine:_nodate} evaluation, each of the system entities were deployed on fresh 64-bit Ubuntu 16.04 distributions on individual Amazon Web Service EC2 virtual server  micro instances one a device with the following specifications: single core Intel(R) Xeon(R) E5-2676 v3 @ 2.40GHz\cite{noauthor_amazon_nodate}. \par

\begin{figure}[h!]
        \centering
        \includegraphics[width=.6\textwidth]{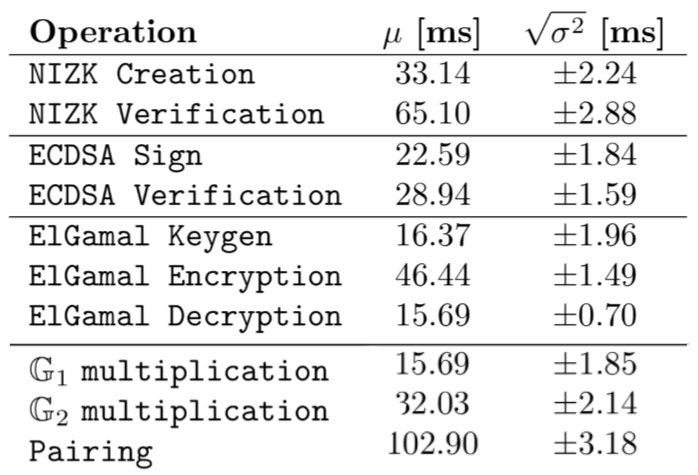}
        \caption{Execution times for cryptographic primitives constructions used in the Tangerine system\cite{stuczynski_tangerine:_nodate}. Measured over 1,000 runs.}
        \label{fig:tang_collage}
    \end{figure}

Figure \ref{fig:tang_collage} shows the results of expected execution time $\mu$ and the corresponding standard deviation $\sqrt{\sigma^2}$ (in $ms$) of the main cryptographic operations and constructs used in the system. 

Naturally, the bilinear pairing operation is the most expensive of the cryptographic operations and constructs, which is the one used for verification. Also, operations in $\mathbb{G}_2$ are twice as expensive as operations in $\mathbb{G}_1$ as expected since the constructs are twice as large. This is why it is important to optimize the system implementation and use the proper fields where necessary in order not to incur extra unneeded costs, even though in theory everything could have been done in $\mathbb{G}_2$.
\par

Even though communication overhead and latency are somewhat negligible in this system and on the order of single/couple kiloByte(s) in size and less than 1 millisecond in delay, they can be optimized further as mentioned in the Tangerine paper\cite{stuczynski_tangerine:_nodate}.

\par
Firstly, the number of authorities present in the system have little to no effect on the client latency as messages are sent simultaneously to all of the authorities in an asynchronous manner. Furthermore, as shown in the Tangerine implementation\cite{stuczynski_tangerine:_nodate}, increasing the number of clients in the system greatly impacts the performance of the system and the perceived delay on the side of the client. This stems from the fact that Node.js is not optimal for heavy computation operations as it is single threaded with an event loop designed for asynchronous, event-driven tasks. However, this is not too much of a concern in this system as functionality and completeness is much more important than latency for an e-petition. In addition, the future goal is to replace the back end of the system with a blockchain based one (as explained in the \nameref{conc} section), which will definitely introduce new bottlenecks in the system.

\clearpage
\section{Conclusions and Future Directions} \label{conc}
This project entails the implementation of the specific application of an e-petition based on the Coconut selective credential disclosure scheme. The main distinctive aspect of this approach is the added feature of unlinkability in addition to anonymity: such that no information about the anonymous petition voter is linkable back to the individual. In other words, the only data which would be leaked \textit{if} at least one authority  maliciously leaks it, is the data concerning who is \textit{eligible} to vote, as opposed to who \textit{actually} voted. This should not happen in the case where all authorities are honest. This censorship resistant property is very useful and sometimes even vital as is the case with the Barcelona referendum \cite{stothard_barcelona:_2018}.

\subsubsection{Signature Petition Scheme}
The signature petition scheme described and implemented in this project is not the most optimal one possible. The way it is done in this project is mainly for simplicity and clarity purposes especially since the main focus was on the Coconut credential system. It can easily be modified or even replaced by a more sophisticated/complex or simplified system based on the specific need.

\subsubsection{Coconut Credential Scheme Template}
During the implementation, it was evident that the Coconut credential scheme could be used for numerous different applications, so the implementation bas branched at the stage where the general Coconut system was ready in order to be used as an open source template.

\subsubsection{Blockchain Based System}
The plan from the start was definitely to implement the system on top of a public, permissionless blockchain platform in order to benefit from its security, immutability, and transparency. Instead of sending information directly to/from the different entities, the blockchain would basically act as an intermediary where the sender publishes what they want to send on the blockchain while the receiver waits until this is done and gets the information off of the blockchain itself. In other words, the blockchain would act as a public bulletin board in which the different parties post whatever they would like to send on to it and the receiving parties get the information directly from the bulletin board. The receiving party would have to keep monitoring the blockchain to check for anything sent to them though, or for efficiency, the sender could notify them to check the blockchain whenever they publish something for example. This is especially valuable in the voting scheme used for the system because it would eliminate the need to trust the authorities and the procedure would be transparent and auditable by everyone.\par

The goal at first was to set up the system to work on top of the permissionless Chainspace\cite{al-bassam_chainspace:_2017}
blockchain because it is perfectly suitable for privacy-preserving applications such as this one. However, unfortunately the Chainspace system is not ready yet at the time of writing. As for Ethereum\cite{buterin_next_nodate}, the Coconut credential system used here can actually be used as an add on to the Open Vote Network\cite{kiayias_smart_2017} (or any other Ethereum based voting system) in order to add the feature of unlinkability to the system.

\subsubsection{Hardware Wallet/Device Integration}
Another future direction which could be a very interesting project would be to integrate hardware wallet/device (such as a Trezor\cite{trezor} or Ledger \cite{ledger}) functionality in order to greatly increase security. The users would have to get issued their signed credential from the authorities on to the hardware wallet/device once at the beginning, to then be able to randomize and use it numerous times in the future. In the case of the e-petition, the user would have to go to the authorities in charge of the petition (government body, university authorities, etc.) to get issued a signed credential from them. Then, whenever they would like to vote for a petition anonymously/unlinkably, they would need to use their hardware wallet/device (which stores their private credential) to create the necessary cryptographic constructs and zero-knowledge proofs in a more secure manner than previously done.

\clearpage
\section{References}
\AtNextBibliography{\small}
\printbibliography[heading=none]

\clearpage
\section{Appendices}

\subsection{Coconut Generic Scheme} \label{coconut_gen}
(As in Coconut paper\cite{sonnino_coconut:_2018})

\begin{description}[leftmargin=1em, labelindent=0em]
\setlength\itemsep{0.5em}
\item[\definition{Setup($1^\lambda$)}{$params$}] Choose a bilinear group $(\mathbb{G}_1,\mathbb{G}_2,\mathbb{G}_T)$ with order $p$, where $p$ is a $\lambda$-bit prime number. Let $g_1, h_1$ be generators of $\mathbb{G}_1$, and $g_2$ a generator of $\mathbb{G}_2$. The system parameters are $params=(\mathbb{G}_1, \mathbb{G}_2, \mathbb{G}_T, p, g_1, g_2, h_1)$. 

\item[\definition{TTPKeyGen($params, t, n$)}{$sk,vk$}] Pick\footnote{This algorithm can be turned into the \textsf{KeyGen} and \textsf{AggKey} algorithms described in Coconut\cite{sonnino_coconut:_2018} using techniques illustrated by Gennaro~\etal~\cite{gennaro1999secure} or Kate~\etal~\cite{cryptoeprint:2012:377}.} two polynomials $v,w$ of degree $t-1$ with coefficients in $\mathbb{F}_p$, and set $(x,y) = (v(0), w(0))$. Issue to each authority $i \in [1, \dots, n]$ a secret key $sk_i = (x_i,y_i) = (v(i), w(i))$, and publish their verification key $vk_i$ = $(g_2,\alpha_i,\beta_i) = (g_2,g_2^{x_i},g_2^{y_i})$.

\item[\definition{IssueCred($m, \phi$)}{$\sigma$}] Credentials issuance is composed of three algorithms:
\begin{description}[leftmargin=1em, labelindent=0em]
\setlength\itemsep{0.5em}
\item \definition{PrepareBlindSign($m, \phi$)}{$d,\Lambda,\phi$} The users generate an \elgamal key-pair $(d, \gamma=g_1^{d})$; pick a random $o\in\mathbb{F}_p$,  compute the commitment $c_m$ and the group element $h\in\mathbb{G}_1$ as follows:
\begin{equation}\nonumber
c_m = g_1^m h_1^o \qquad{\rm and}\qquad h = \hashtopoint(c_m)
\end{equation} 
Pick a random $k \in \mathbb{F}_p$ and compute an \elgamal encryption of $m$ as below:
\begin{equation}\nonumber
c = Enc(h^m)=(g_1^k,\gamma^k h^m)
\end{equation}
Output $(d, \Lambda=(\gamma, c_m, c, \pi_{s}), \phi)$, where $\phi$ is an application-specific predicate satisfied by $m$, and $\pi_{s}$ is defined by:
\begin{eqnarray}\nonumber
\pi_{s} &=& {\rm NIZK}\{(d, m, o, k): \gamma = g_1^d \;\land\; c_m=g_1^mh_1^o\\ \nonumber
 && \land\; c = (g_1^k,\gamma^k h^m) \;\land\;  \phi(m)=1\}
 \end{eqnarray}

\item \definition{BlindSign($sk_i, \Lambda, \phi$)}{$\tilde{\sigma}_i$} The authority $i$ parses $\Lambda=(\gamma, c_m, c, \pi_{s})$, $sk_i=(x,y)$, and $c=(a,b)$. Recompute $h = \hashtopoint(c_m)$. Verify the proof  $\pi_{s}$ using $\gamma$, $c_m$ and $\phi$; if the proof is valid, build $\tilde{c}=(a^y,h^xb^y)$ and output $\tilde{\sigma}_i = (h, \tilde{c})$; otherwise output $\perp$ and stop the protocol.

\item \definition{Unblind($\tilde{\sigma}_i, d$)}{$\sigma_i$} The users parse $\tilde{\sigma}_i=(h, \tilde{c})$ and $\tilde{c}=(\tilde{a},\tilde{b})$; compute $\sigma_i = (h,\tilde{b}(\tilde{a})^{-d})$. Output $\sigma_i$.
 \end{description}
 
\item[\definition{AggCred($\sigma_1, \dots, \sigma_t$)}{$\sigma$}] Parse each $\sigma_i$ as $(h,s_i)$ for $i \in [1, \dots, t]$. Output $(h,\prod^t_{i=1} s_i^{l_i})$, where $l$ is the Lagrange coefficient:
\begin{equation}\nonumber
l_i = \left[\prod^t_{i=1, j\neq i} (0-j)\right] \left[\prod^t_{i=1,s j\neq i} (i-j)\right]^{-1} \;{\rm mod}\; p
\end{equation}

\item[\definition{ProveCred($vk, m, \sigma, \phi'$)}{$\Theta,\phi'$}] Parse $\sigma=(h,s)$ and $vk=(g_2,\alpha,\beta)$. Pick at random $r',r \in \mathbb{F}_p^2$; set $\sigma'=(h',s')=(h^{r'},s^{r'})$; build $\kappa = \alpha\beta^m g_2^r$ and $\nu=\left(h'\right)^r$. Output $(\Theta=(\kappa, \nu, \sigma',\pi_v),\phi')$, where $\phi'$ is an application-specific predicate satisfied by $m$, and $\pi_v$ is:
\begin{equation}\nonumber
    \pi_v={\rm NIZK}\{(m,r): \kappa=\alpha\beta^m g_2^r \ \land \ \nu=\left(h'\right)^r \ \land \  \phi'(m)=1\} 
\end{equation}

\item[\definition{VerifyCred($vk, \Theta, \phi'$)}{$true/false$}] Parse $\Theta = (\kappa, \nu, \sigma',\pi_v)$ and $\sigma'=(h',s')$; verify $\pi_v$ using $vk$ and $\phi'$. Output $true$ if the proof verifies, $h'\neq1$ and $e(h',\kappa)=e(s'\nu,g_2)$; otherwise output $false$.
\end{description}

\end{document}